# Tunable coupling-induced resonance splitting in self-coupled Silicon ring cavity with robust spectral characteristics

Awanish Pandey and Shankar Kumar Selvaraja

Photonics Research Laboratory, Centre for Nano Science and Engineering,
Indian Institute of Science, Bengaluru, 560012
India

To whom correspondence should be addressed; E-mail: awanish@cense.iisc.ernet.in

We propose and demonstrate a self-coupled micro ring resonator for resonance splitting by mutual mode coupling of cavity mode and counter-propagating mode in Silicon-on-Insulator platform The resonator is constructed with a self-coupling region that can excite counter-propagating mode. We experimentally study the effect of self-coupling on the resonance splitting, resonance extinction, and quality-factor evolution and stability. Based on the coupling, we achieve 72% of FSR splitting for a cavity with FSR 2.1 nm with ¡ 5% variation in the cavity quality factor. The self-coupled resonance splitting shows highly robust spectral characteristic that can be exploited for sensing and optical signal processing.

Advances in Silicon photonics (SiP) has provided tremendous impetus to achieve the next generation high-speed short and long reach communication. It provides the advantages of optical transparency and large-bandwidth along with CMOS compatibility to deliver compact energy-efficient circuits[1]. Moreover, improvements in the fabrication technology has enabled low-loss waveguides and highly-uniform device response that was once considered a challenge in adapting SiP for practical applications [2].

Micro-Ring Resonators (MRRs) are one of the fundamental building blocks of SiP integrated circuits. Si MRRs have found applications in areas such as sensors [3], modulators [4] and filters [5]. A simple MRR consist of a ring waveguide evanescent-coupled to one or two straight waveguides. Typically, MRRs have only forward propagating cavity modes at resonant wavelengths. However, in practice, any non-ideality in the resonator can result in the generation of counter-propagating modes in the cavity. The two primary sources of such non-idealities are sidewall roughness in the waveguides and non-unidirectional coupling between the bus and the ring waveguide. The sidewall roughness act as scattering points that redistributes the energy into counter-propagating mode, while coupling region between a ring and the straight waveguide can excite contra-directional modes because of reflections caused due to mode mismatch between the straight bus waveguide and the cavity mode[6, 7]. Such counter-propagating modes result in resonance splitting and are referred to as Autler-Townes splitting (ATS) of resonance [8]. ATS can create a transparency window at the resonator output which is undesirable for most applications relying on cavity modes [9].

Furthermore, the inability to predict or control the shape and symmetry of the splitting makes it challenging to design and fabricate reliable devices based on MRR. Another drawback of such splitting is power leakage that may cause cross-talk and reduces the power at the desired output port rendering the configuration inefficient [10]. Several methods have been proposed to suppress the splitting such as lowering the Q-factor, making smooth waveguide sidewalls and operating in TM mode instead of TE to reduce the interaction of cavity mode with the sidewalls [11]. Such schemes require highly accurate atomically smooth device fabrication, which is challenging even with advanced CMOS fabrication technology.

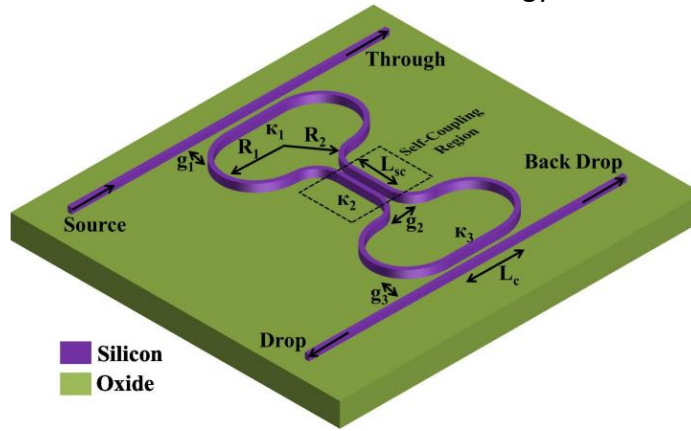

Figure 1: Schematic of SCMRR. Counter propagating modes are generated in the self-coupling region defined by a directional coupler with length $L_{sc}$ and gap $g_2$.

A potential solution to unpredictable nature of the resonance splitting is to create an intentional and tunable perturbation in the resonator that can effectively control the generation of counter propagating waves. Such techniques yield interesting applications such as photonic links [12] and signal processing [13]. The generation of such counter-propagating modes can be achieved either by embedded coupled cavity system, corrugating the cavity sidewall or employing quasi-grating [13, 14, 15]. However, the embedded coupled cavity system requires high precision of cavity dimensions whereas the quasi grating and corrugated side-wall structures suffer from scattering losses that would compromise the quality factor of the cavity. Thus, it would be advantageous to design a resonating structure that provides tunable, fabrication tolerant, low-loss and a reliable way of generating counter propagating mode in the cavity.

In this letter, we present a Self-Coupled Micro-Ring Resonator (SCMRR) configuration that can create clockwise and counter-clockwise propagating modes in the cavity by tuning the self-coupling strength. A schematic of the SCMRR is shown in Fig. 1, it is constructed as a hour-glass loop with a central self-coupling region and two straight bus waveguides. The SCMRR is coupled to the bus waveguides with a coupling coefficient of $\kappa_1$ and $\kappa_3$, while the self coupling is created by the central self-coupling coefficient $\kappa_2$. The bus waveguides on either sides of the SCMRR serves as input and output waveguides. A SCMRR has four port; Source (SP), Through (TP), Drop (DP) and Back-Drop (BDP) or Add port as illustrated in Fig. 1.

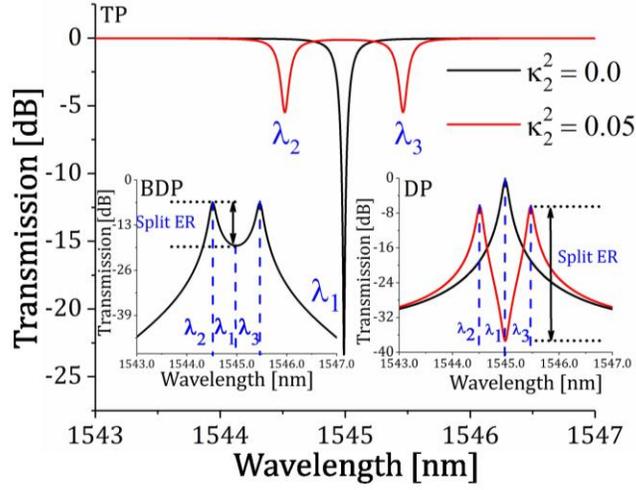

Figure 2: Simulated transmission spectra of a SCMRR for two different self-coupling coefficients ($\kappa_2 = 0$ and $\kappa_2 = 0.05$).

Fig. 2 shows the simulated response of a SCMRR using coupling coefficient matrix method [16]. For simplicity, bus-to-SCMRR coupling is kept identical; $\kappa_1 = \kappa_3$ for all the simulations. Furthermore, we have considered SOI waveguide of 450×220 nm cross-section with a propagation loss of 2.5 dB/cm. When $\kappa_2 = 0$ and $\kappa_1 = \kappa_3 = 5\%$, the SCMRR behaves as a simple MRR that resonates at $\lambda_1$ (Eq. 1) creating a transmission dip at the TP and transmission peak at the DP while no power is present at the BDP (inset Fig. 2).

$$n_{eff} * 2\pi(2R_1 + R_2) + 2L_{sc} + 2L_c = m.\lambda \qquad (1)$$

where $n_{eff}$ refers to the effective refractive index of propagating mode in the SCMRR and $m$ refers to the resonance order. The inset in Fig. 2 shows the DP and BDP response.

When $\kappa_2 \neq 0$, counter propagating mode is generated that results in the symmetric resonance splitting of $\lambda_1$; $\lambda_2$ and $\lambda_3$. Similarly, in the DP, the transmission peak splits creating a large insertion loss (IL) at $\lambda_1$. The splitting is generated by lifting of frequency degeneracy caused by the interference of counter-propagating degenerate modes travelling at the resonance wavelengths. The lifting of degeneracy results in symmetric doublets in wavelength because of strong inter-resonance coupling [17]. In the DP, however, the IL increases due to splitting. The IL of the resonance wavelength at the DP increases from 0.4dB to 5dB and in BDP, an IL of 5dB is observed, where no light was initially present. The split-ER between the symmetric resonances centred round $\lambda_1$ at BDP and DP is 18dB and 37dB respectively. This large difference of ER at minima for both the ports arises from the energy in the counter-propagating mode. Weaker coupling at the SCR results in a weak perturbation in the resonator, a 3dB splitting in the SCR would result in an identical spectral response with equal split-ER in the DP and BDP. A notable characteristics of such mode splitting scheme is the considerable reduction in the TP ER. The ER at the resonance wavelength drops from 24dB to 5dB due to distribution of power between the two counter-propagating degenerate modes. The ER can be improved by increasing $\kappa_1$ and $\kappa_3$ as it drives the split resonances in different coupling regimes. [18].

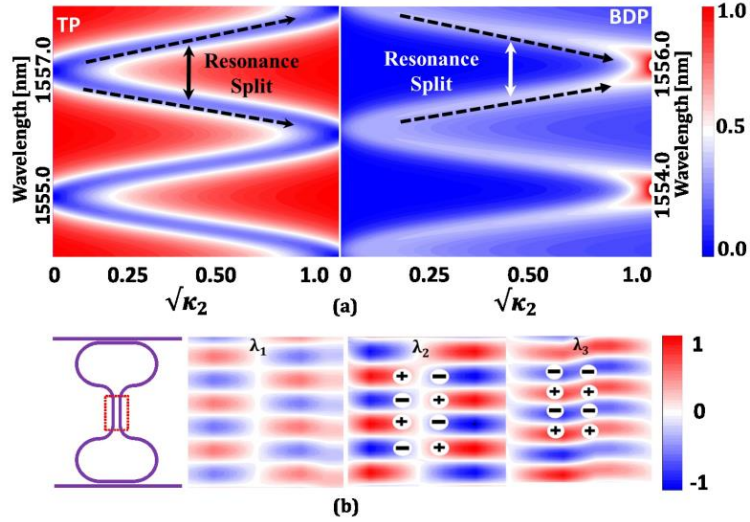

Figure 3: (a) shows evolution of splitting for TP and BDP respectively with $\kappa_2$ whereas (b) shows field distribution in SCMRR for three cases, at $\lambda_1$ when $\kappa_2 = 0$ and at $\lambda_2$ and $\lambda_3$ when $\kappa_2 \neq 0$.

Fig. 3(a) illustrates the effect of $\kappa_2$ on the TP and BDP spectral response of SCMRR. At TP, when sqrt($\kappa_2$) = 0, two minima occur at 1555 nm and 1557 nm corresponding to unperturbed cavity resonance. As $\kappa_2$ becomes non-zero, split is observed which progressively increases and reaches at $FSR/2$ at sqrt($\kappa_2$) = 0.5. Furthermore as $\kappa_2$ extends towards one, the splits from neighbouring resonances overlap and results in a spectrum with a phase shift of $\pi$ compared to the initial condition; $\kappa_2 = 0$. In the case of BDP, no transmission is observed with zero self-coupling. However, the transmission starts increasing for non-zero self-coupling and when $\kappa_2 = 1$, the entire optical power is redirected to BDP and no power is found in DP. The maxima at BD, at $\kappa_2 = 1$, will appear at 1554 nm and 1556 nm which is $FSR/2$ shifted version of DP at $\kappa_2 = 0$. Since the behaviour of the SCMRR strongly depends on the $\kappa_2$ in SCR, a 3D FDTD simulation was done on a configuration with $\kappa_2 = 0$ and $\kappa_2 \neq 0$ to study the effect of SCR on lifting of degeneracy (Fig. 3(b)). When $\kappa_2 = 0$, obviously there is no interaction between the SCR waveguides. However, when $\kappa_2 \neq 0$ the cavity resonance splits and interestingly the field distribution in the SCR shows symmetric and anti-symmetric coupling at the split resonances $\lambda_2$ and $\lambda_3$ respectively [19].

The fabrication of SCMRR was done on a SOI wafer with 220 nm Silicon device layer on top of 2000 nm buried-Oxide. The devices were patterned using e-beam lithography and dry etch process. For e-beam pattering, ma-N 2401 resist was used followed by 220 nm of Si dry etching by inductively coupled plasma (ICP-RIE) using $SF_6$ and $C_4F_8$ gas chemistry. To couple light in and out of the device, grating fiber-chip couplers were fabricated also using e-beam process with PMMA resist and subsequently etched 70 nm into Si using $SF_6$ and $CHF_3$ gas chemistry. Fig. 4 shows SEM image of a SCMRR. Devices with varying coupling ($\kappa_2$) were fabricated and optically characterised.

Optical transmission characterisation was performed using a tunable laser (Keysight 81960A) and a powermeter in the C&L band. As mentioned earlier, light is coupled in and out of the device through the grating couplers. At the input side, polarisation paddles were used to control the input polarisation while the output is fed into the powermeter.

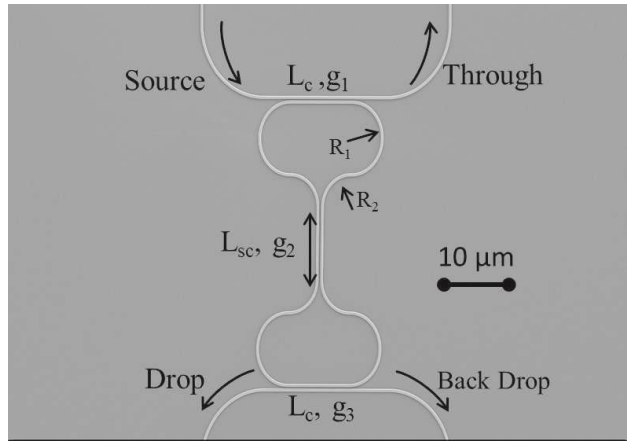

Figure 4: SEM image of fabricated SCMRR.

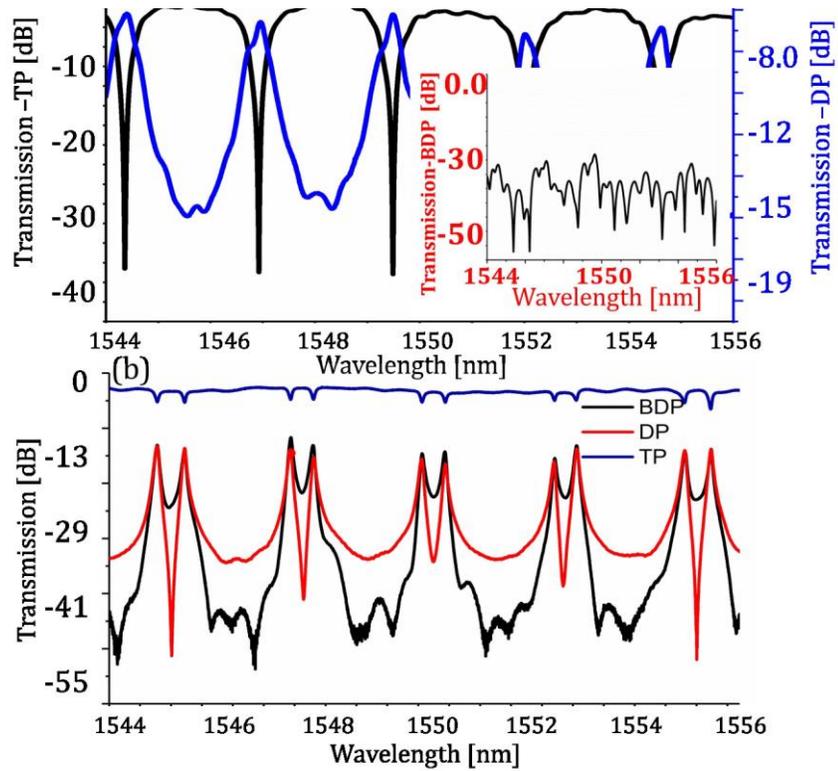

Figure 5: Transmission spectra of all the three ports in SCMRR with (a) $\kappa_2 = 0$ and (b) $\kappa_2 \neq 0$.

Fig. 5(a) shows a typical transmission spectral response of a SCMRR with $\kappa_2 = 0$ and Fig 5(b) shows the transmission for $\kappa_2 = 2.4\%$ with 3% bus-to-ring coupling; $\kappa_1$ and $\kappa_2$. Zero self-coupling results in conventional MRR response with very low-power < 30dB at resonance appearing at BDP. However, with non-zero central coupling, the resonance splits and considerable amount of optical power appears at BDP as shown in Fig. 5(b). As discussed earlier, the resonance splitting is symmetric in nature with different split-ER for DP and BDP. The insertion loss at the DP increases from 3dB to 12.3 dB due to splitting. Since the energy is distributed between the clockwise and counter-clockwise propagating modes, the energy is split between the DP and BDP, resulting in higher IL at the DP. Though the IL of the DP and BDP varied between 11-13 dB over the different resonance wavelengths, the split-ER, that is the extinction between the split resonances shows a clear difference. The split-ER of DP varies from 22 dB to 36 dB whereas for the BDP it is 8 dB to 11 dB. Similarly, the TP extinction decreases with self coupling $\kappa_2$. Fig. 6 captures the effect of central coupling on the TP extinction at the resonance.

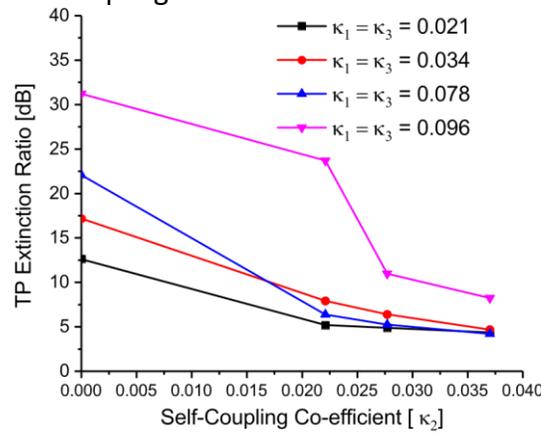

Figure 6: Measured Through Port ER as a function of self coupling $\kappa_2$ with different bus-to-ring coupling coefficients $\kappa_1$ and $\kappa_3$.

The coupling between the bus and the ring is kept symmetric ($\kappa_1 = \kappa_3$) and $\kappa_2$ is varied from 0% to 3.5%. With an increase in $\kappa_2$ the TP extinction reduces consistently and saturates. As observed in Fig. 6 with increase in the bus-to-ring coupling ($\kappa_1$ and $\kappa_3$) the TP extinction could be improved.

The SCMRR with $\kappa_2 = 0$ is critically coupled, however, as the value of $\kappa_2$ increases the effective loss in the cavity also increases pushing the resonator to operate in an under-coupled regime which decreases the TP ER. For a higher bus-to-ring coupling ($\kappa_1$=9.6%) the TP ER drops from 32 dB to 8 dB while at lower coupling ($\kappa_1$=2.1%) it drops from 13 dB to 3 dB. However, at the DP and BDP the split-ER increases with increasing $\kappa_2$. Fig. 7a and Fig. 7b illustrates the evolution of the resonance splitting as a function of self-coupling coefficient $\kappa_2$. The splits consistently increases with $\kappa_2$ for both DP and BDP and goes upto 1.52 nm for $\kappa_2 = 0.03$ which is around 72% of the unperturbed cavity FSR (2.1 nm). Furthermore, the quality factor of the resonances 8(a) and 3dB linewidth 8(b) in all the three ports are analysed. The quality factor of the resonances in all the ports were highly stable, which shows the uniformity in the fabrication process and device robustness. When $\kappa_2$ becomes non-zero, the quality factor at the DP marginally reduces and stabilises at around 13,403± 349 and 13,308± 550 for BDP and DP respectively. The initial drop in the DP Q-factor can be attributed to the additional loss incurred in the self-coupling directional coupler. The TP Q-factor, however, remains stage irrespective of $\kappa_2$ at 12,600 ± 282.

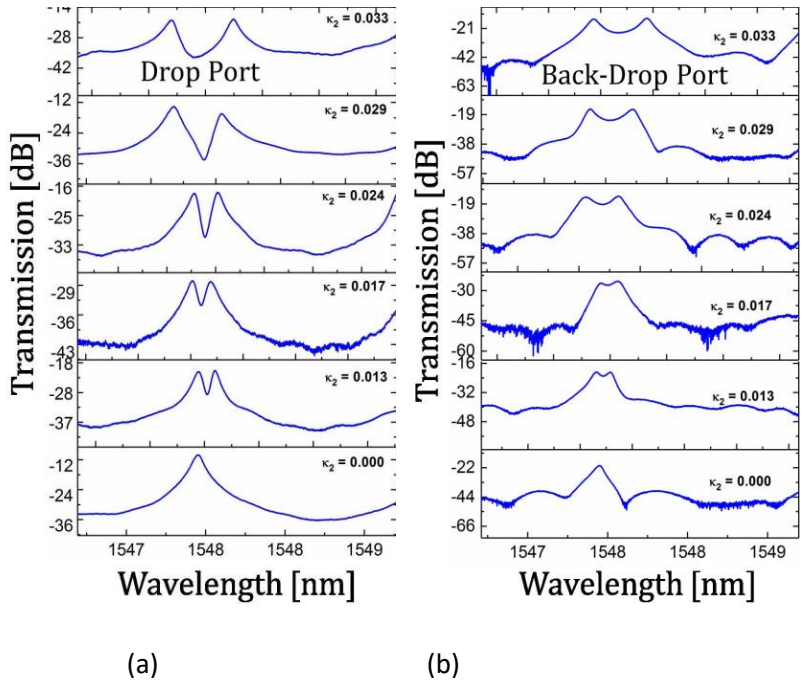

(a) (b)

Figure 7: Evolution of drop and back-drop port response as function of $\kappa_2$.

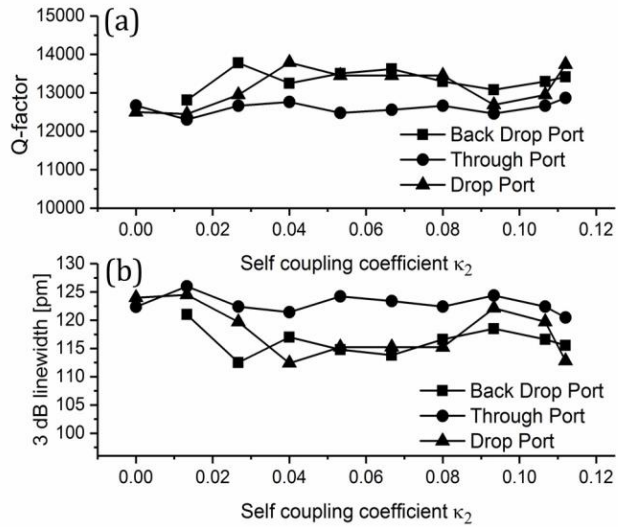

Figure 8: Measured (a) Q-factor and (b) FWHM of the split resonances at all the output ports of SCMRR.

In conclusion, we have presented and experimentally demonstrated a self-coupled microring resonator configuration that can create and control resonance splitting in ring resonator type systems. The self-coupling introduces a controlled splitting in the resonance cavity that can generate counter-propagating degenerate mode in the cavity. The design offers duplicated spectral response at two output ports. The simulated response and characteristics of the resonance are validated experimentally. The proposed SCMRR based photonic circuits can be used as a wavelength filter, cavity mode-hybridisation in photonic molecules, on-chip non-linear, sensing and RF-photonics signal processing.

The authors thank the staff of the National Nano-Fabrication Centr (NNFC) and the Micro and Nano Characterization Facility (MNCF) at the Indian Institute of Science, Bengaluru for their assistance. Authors AP is grateful to MeitY, Government of India for providing his Visvesvaraya fellowship and SKS for his Visvesvaraya faculty fellowship.